\documentclass[11pt,twoside]{article}
\usepackage{asp2010}

\resetcounters

\begin{document}

\title{Searching of New Emission-Line Stars using the\\ Astroinformatics
Approach}

\author{Petr \v{S}koda and Jaroslav V\'{a}\v{z}n\'{y}
\affil{Astronomical Institute of the Academy of Sciences, Fri\v{c}ova 298,\\
251\,65 Ond\v{r}ejov, Czech Republic}
}

\begin{abstract}
Using data mining techniques applied on emission line characteristics  of Be
stars spectra  we attempted  to  find new Be stars candidates in SDSS SEGUE
survey.
The mid-resolution spectra of confirmed Be stars obtained from VO-compatible
archive of Ond\v{r}ejov observatory 2m telescope  were transformed to the
spectral resolution of SDSS  and important characteristics of emission
line profiles were estimated, to be used as a training base of supervised
learning methods. The obtained knowledge base of the characteristic shapes and
sizes of Be emission lines was finally used to identify new potential
candidates in SDSS spectral survey. The several newly found
Be stars candidates justify our  approach and approve Astroinformatics as a
viable research methodology.
\end{abstract}

\section{Introduction}

Current data deluge in astronomy requires applying data mining techniques to
extract new information about the physical nature of celestial objects. The
possibility of cross-matching several surveys via Virtual Observatory
protocols may play a key role in future discoveries. Data mining of large
collections of spectra seems to be one of promising as well as challenging
topics.  We have focused on obtaining new candidates of H$\alpha$ emission
stars using supervised data mining method of Decision Trees on almost 200,000
spectra in SDSS SEGUE \citep{yanny2009segue} spectral survey.

\section{Data Sources}
The spectra obtained with coud\`e spectrograph of Ond\v{r}ejov
Observatory 2m telescope were used as a training sample.
Using Virtual Observatory protocol SSA the spectra
from Ond\v{r}ejov 2m telescope archive server were acquired based on the list of justified Be stars obtained from other studies. After required pre-processing of spectra the vectors of spectral line shape parameters
characterizing the typical Be star H$_\alpha$ emission line were obtained and
subjected to data mining process.

As a testing sample the spectra from project SEGUE of SDSS were
selected. This contains 178314 spectra in DR7. A simple SQL query was
used to generate the list of URL links for individual FITS
files  downloaded afterwards.

\section{Spectra Preprocessing}
Before the real data mining process can be started all input data have to be converted to the common representation in the terms of the same flux units, number of pixels or spectral resolution power.

\subsection{Degradation of Spectral Resolution}

Spectra from Ond\v{r}ejov Observatory have higher spectral resolution
than SDSS, therefore the degradation of spectral resolution was
applied on them followed by re-binning to the same number of pixels as the
SDSS. So we obtained the training set of Ond\v{r}ejov Be stars spectra looking
similar to  SDSS spectra.  
For that purpose convolution in discrete form was used

\begin{equation}
  \label{eq:discreteConvolution}
  (f * g)[n]\ \stackrel{\mathrm{def}}{=}\ \sum_{m=-\infty}^{\infty} f[m]\, g[n - m]
\end{equation}

\noindent An example of this process applied on spectra of Be star 4 Her is on the
Fig.~1. The top figure shows Gaussian function used for convolution
with the spectrum, followed by the original spectrum, then there is a
spectrum after convolution with the Gaussian profile. The last is the
final spectrum after re-binning.

\begin{figure}[!ht]
\plotfiddle{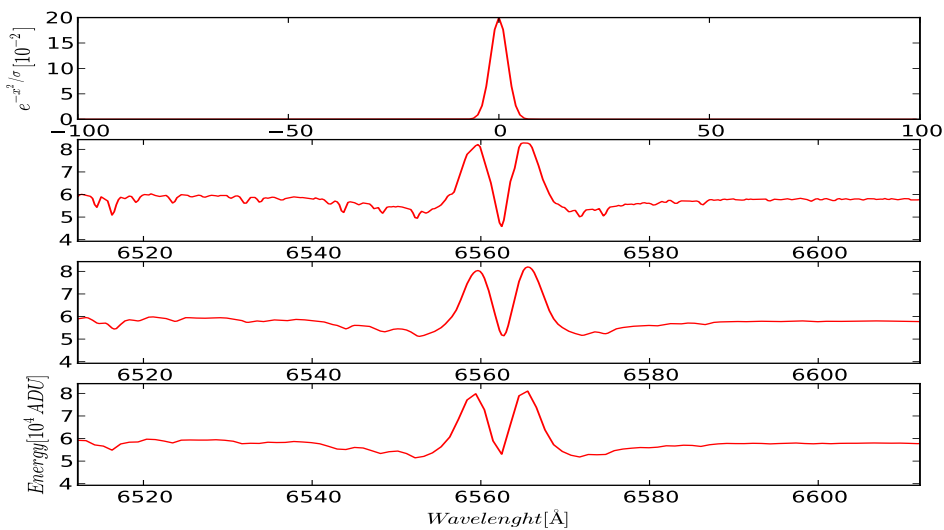}{7cm}{0}{70}{50}{-200}{0}
\caption{The convolution with SDSS instrumental profile and re-binning applied on Ond\v{r}ejov spectra}
\end{figure}

\subsection{Continuum Normalization} 

As spectra from SDSS are absolute flux calibrated, while the Ond\v{r}ejov
spectra state intensity only in ADUs,  the rectification of spectra was
performed to be able to compare the size and shapes of spectral line profiles.
The spectra were divided by the robust linear function roughly representing
its (pseudo)continuum. To  ensure the compatibility for data mining process only
the spectral range covered by Ond\v{r}ejov spectra ($6300-6800\AA$) was
considered in fitting the line in SDSS spectra.

\begin{figure}[!ht]
\plotfiddle{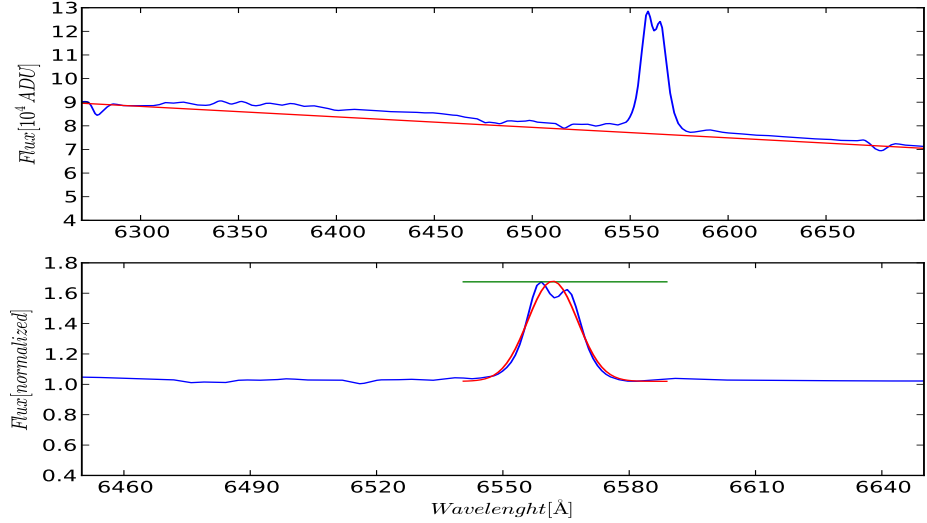}{7cm}{0}{70}{50}{-200}{0}
\caption{The normalized spectrum of Be star 60 Cyg with fitted Gaussian curve} 
\end{figure}

\noindent The top panel of Fig.~2 
      depicts the continuum fit. The bottom figure shows the region
      (width of the green line) used for extraction. The position of
      the line corresponds to the maximum value in the region of
      $50\,\textrm{\AA}$. The Gaussian fit is in red. Although the fit
      is almost perfect, this approach fails to get characteristic
      double peak of the emission line.

\section{Spectral Line Parameters}

As parameters for data mining process characteristic shape of
H$\alpha$ line were extracted from the spectra. Three parameters were
finally selected. The height and the width of the H$\alpha$ emission
line and median absolute deviation as a characterization of the noise
level in the spectrum.

\noindent{\em The height of the H$\alpha$ line}

\noindent The maximum value in the region of $50\,\textrm{\AA}$ around H$\alpha$
above the linear fit was extracted from the spectrum.

\noindent{\em The noise level of the spectrum}

\noindent The noise in the spectrum contributes to the characteristics of the
spectral lines. As an estimator of the noise level the median
absolute deviation was used, defined as:

\begin{equation}
  \textrm{mad} = \textrm{median}_{i}\left(\ \left| X_{i} -
      \textrm{median}_{j} (X_{j}) \right|\ \right)
\end{equation}

\noindent{\em The width of the H$\alpha$ line}
\noindent The Gaussian function:

\begin{equation}
  \label{eq:gauss2}
  f(x) =  1 + e^{- { \frac{(\lambda-\lambda_0)^2 }{S^2} } }
\end{equation}

\noindent was fitted to the profile of H$\alpha$ spectral line. 
The robust estimators \citep{launer1979robustness} of line center and width were computed which minimize the
sum of squares

\begin{eqnarray}
  x_0  &= \textrm{median}( w_jx_j ) /\ \sum{w_i}, \\
  S  &= \textrm{mad}( \lambda_i - \lambda_0 ) /\ \sum{w_i}.
\end{eqnarray}

\section{Data Mining}

The decision tree based classification was performed using Weka software with
algorithm J48, which is the free implementation of algorithm C4.5. The training
set had 173 and testing set 178314 items.

\section{Conclusions}
\begin{figure}[!htb]
\plotfiddle{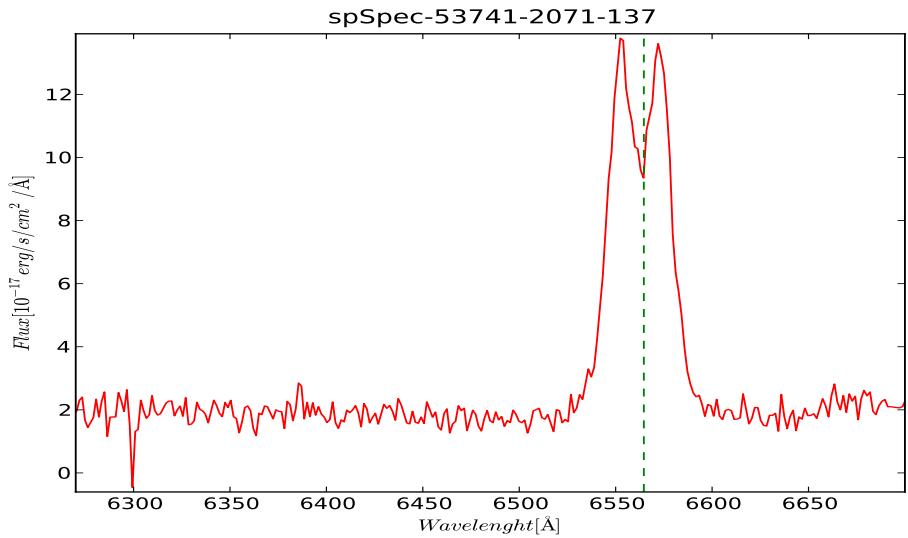}{6.2cm}{0}{70}{50}{-200}{20}
\caption{The example of candidate Be star found in SDSS SEGUE survey}
\end{figure}

\noindent The classifier has identified 1110 Be stars
candidates in SEGUE, however most of them are probably of different nature
(e.g. AGNs, young stellar objects or reduction artifacts). Nevertheless, there
are as well several highly probable  Be stars like the one on Fig.~3.

\acknowledgements The Astronomical Institute of the Academy of Sciences of the
Czech Republic is supported by project AV0Z10030501. This work is part of
diploma thesis \citep{vazny}.

\bibliographystyle{asp2010}
\bibliography{P141}

\end{document}